\documentclass[aps,prl,twocolumn,tightenlines,showpacs]{revtex4}
\usepackage{epsfig,rotating,amsmath}
\usepackage{multirow}

\begin{document}

\title{Nuclear Matrix Elements for Tests of Local  Lorentz Invariance Violation}
\author{B. A. Brown$^{1}$, G. F. Bertsch$^{2}$, L. M. Robledo$^{3}$, M. V.
Romalis$^{4}$ and V. Zelevinsky$^{1}$}
\affiliation{$^{1}$ Department of Physics and Astronomy and National Superconducting
Cyclotron Laboratory, Michigan State University, East Lansing, Michigan
48824-1321, USA}
\affiliation{$^{2}$ Institute for Nuclear Theory and Department of Physics, Box 351560,
University of Washington, Seattle, Washington 98915, USA}
\affiliation{$^{3}$ Departamento de Fisica Teorica, Modulo 15, Universidad Autonoma de
Madrid, E-28049 Madrid, Spain}
\affiliation{$^{4}$ Department of Physics, Princeton University, Princeton, New Jersey
08544, USA}

\begin{abstract}
The nuclear matrix elements for the
momentum quadrupole operator are important for the
interpretation of precision atomic physics experiments
that search for violations of local Lorentz and CPT symmetry and for new
spin-dependent forces. We use the configuration-interaction nuclear
shell model and self-consistent mean field theory to calculate these
matrix elements in $^{21}$Ne,  $^{23}$Na, $^{131}$Xe, $^{173}$Yb and $^{201}$Hg.
These are the first microscopic calculations of the nuclear matrix elements
for the momentum quadrupole tensor that go beyond the single-particle estimate.
We show that they are
strongly suppressed by the many-body correlations, in contrast to the
well known enhancement of the
spatial quadrupole nuclear matrix elements.
\end{abstract}


\maketitle

Several types of precision low energy tests of the Standard Model
use nuclear-spin-polarized atoms to achieve very high sensitivity by
relying on long nuclear spin coherence times that are possible with
atoms in $  ^{1}S_{0}  $ ground state, such as $^{3}$He, $^{21}$Ne,
$^{129}$Xe, $^{131}$Xe, $^{173}$Yb, $^{199}$Hg and $^{201}$Hg. Such tests
include searches for violation of local Lorentz and CPT symmetry
\cite{hg201}, \cite{Walsworth}, \cite{Brown}, \cite{Smiciklas},
and for new spin-dependent forces
mediated by light particles,  such as an axion
\cite{Venema}, \cite{Youdin},
\cite{Vasilakis}, \cite{Bulatowicz}, \cite{Tullney}, \cite{Hunter}.

The interpretation and comparison of these experiments requires knowledge of
nuclear matrix elements required for new interactions beyond the Standard
Model. A number of simple models have been used to estimate the relevant
nuclear matrix elements \cite{Kostclock}, \cite{Flambaum}, \cite{Kimball}, \cite{flam},
\cite{flamc}, \cite{lackenby}.
but few detailed
nuclear structure calculations have been performed so far for this purpose.
This can be contrasted with a large number of nuclear structure calculations
performed to estimate the scattering cross-sections for dark matter
particles \cite{Revmat}, \cite{Xespin}, \cite{an2014}
and rates for neutrinoless double-beta decay
\cite{doublebeta}.

The nuclear matrix elements relevant in searches for
local Lorentz invariance violation (LLIV) within the Standard Model Extension (SME) are
derived in \cite{Kostclock}. Here we focus on matrix elements that
generate couplings to CPT-odd $  b_{\mu }  $ and CPT-even $  c_{\mu \nu }  $
terms in the SME Lagrangian for fermions:
$$
\mathcal{L}=\frac{1}{2}i\overline{\psi }(\gamma _{\nu }+c_{\mu \nu}
\gamma ^{\mu})\overleftrightarrow{\partial }^{\nu }\psi
-\overline{\psi }(m+b_{\mu }\gamma _{5}\gamma ^{\mu })\psi.       \eqno({1})
$$
For non-relativistic nucleon motion they generate an energy shift
$$
\mathcal{H}=-2b_{j}S_{j}-(c_{jk}+c_{00}\delta _{jk}/2)p_{j}p_{k}/m,       \eqno({2})
$$
where $  S_{j}  $ is the spin operator, $  p_{j}  $ is the momentum
operator, and $  m  $ is the mass of the fermion. Traditionally, LLIV
effects and spin-dependent forces have been analyzed
separately at the level of neutrons and protons under the assumption
that they are independent. This provides a way to roughly classify
the experiments without making assumptions about a microscopic
theory that would likely generate comparable effects in neutrons and
protons. For particles that are on average at rest, only the
spherical rank-2 components of the tensor $  p_{i}p_{j}  $ give a finite
energy shift. Using Wigner-Eckart theorem, they can be expressed in
terms of the matrix elements of the momentum quadrupole tensor operator
$  {\hat M} = 2 p_{z}^{2} -p_{x}^{2} -p_{y}^{2}  $,
$$
M=\left\langle I,I\left\vert
{\hat M}\right\vert I,I\right\rangle=
\left\langle I,I\left\vert
2p_{z}^{2}-p_{x}^{2}-p_{y}^{2} \right\vert I,I\right\rangle,       \eqno({3})
$$
for a nucleus with spin $  I  $ and its projection $  I_{z}=I  $.
In the nucleus there are two components for this,
proton, $  M_{p}  $, and neutron, $  M_{n}  $.
The best current limits on LLIV
effects currently come from the
quadrupole momentum matrix element in the nucleus $^{21}$Ne \cite{Smiciklas}.
The calculations for $^{21}$Ne \cite{Smiciklas} were based
on a simple single-particle estimate for the
odd valence neutron. Flambaum et al. \cite{flam},
\cite{lackenby} have presented a model where
momentum quadrupole moment ($  M  $) is related to the experimental
spatial quadrupole moment ($  Q  $)
$$
Q=\left\langle I,I\left\vert
{\hat Q}\right\vert I,I\right\rangle=
\left\langle I,I\left\vert
2z^{2}-x^{2}-y^{2} \right\vert I,I\right\rangle,       \eqno({4})
$$
with two components $  Q_{p}  $ and Q$_{n}  .  $

In addition to $^{21}$Ne (odd nucleon), in this letter we consider four other nuclei that all have 
$  I \geq 3/2  $
that is required for the tensor matrix elements to be nonzero.
Three heavy nuclei $^{133}$Cs (odd proton), $^{173}$Yb (odd neutron), and $^{201}$Hg (odd neutron) 
are
used widely for atomic NMR studies.
$^{133}$Cs can be used in an alkali-metal co-magnetometer, using techniques similar to \cite{cs133}.
$^{173}$Yb can be used using an optical dipole trap \cite{yb173}.
LLIV for $^{201}$Hg was studied in \cite{hg201}.
For consistency we also consider the odd-proton $  sd  $ shell nucleus $^{23}$Na.
We present  self-consistent mean-field model (SCMF)
calculations for all of these nuclei
that are consistent with the experimental
$  Q  $ for protons. For heavy nuclei the momentum matrix elements $  M  $ are close to
zero within the theoretical uncertainty. We give
a simple explanation for this result. These results are inconsistent
with previous calculations \cite{flam}, \cite{lackenby}. For $^{21}$Ne and $^{23}$Na we compare the
SCMF results to those from the configuration interaction (CI) shell model. Within the CI model
it is essential to include core polarization that reduces the momentum matrix elements
compared to those obtained in the $  sd  $ shell valence space. The consistency of the SCMF
and CI results for $^{21}$Ne and $^{23}$Na suggest small but robust non-zero values for the
tensor momentum matrix elements.

\begin{figure}
\includegraphics[scale=0.6]{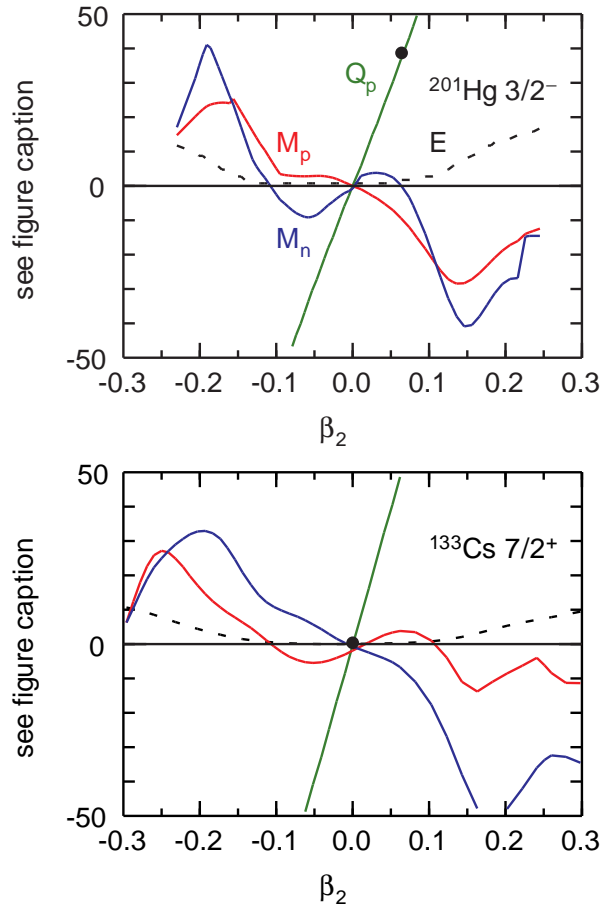}
\caption{Results of the SCMF calculations for $^{133}$Cs and $^{201}$Hg.
Four curves are shown as a function of the constrained $\beta_{2}$ value.
The dashed line labeled $  E  $ is the SCMF energy
in units of MeV relative to its
minimum. The green line labeled $  Q_{p}  $ is the charge quadrupole moment in units
of e fm$^{2}$.
The blue line labeled $  M_{n}  $
is the neutron momentum quadrupole moment in units of $  m  $ MeV,
where $  m  $ is the nucleon mass.
The blue line labeled $  M_{p}  $
is the proton momentum quadrupole moment in units of $  m  $ MeV.
The experimental charge quadrupole moment \cite{nnds} is shown by the black circle
on the green line.
}
\end{figure}

\begin{figure}
\includegraphics[scale=0.6]{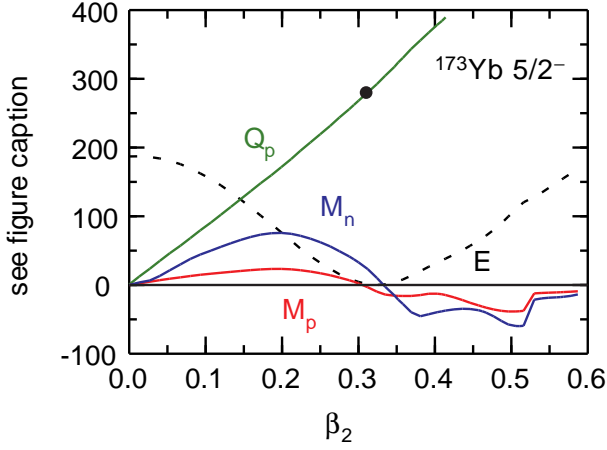}
\caption{Results of the SCMF calculations for $^{173}$Yb.
The labels and units are the same as Fig. 1, except that $  E  $ has
been multiplied by 10.
}
\end{figure}

\begin{figure}
\includegraphics[scale=0.6]{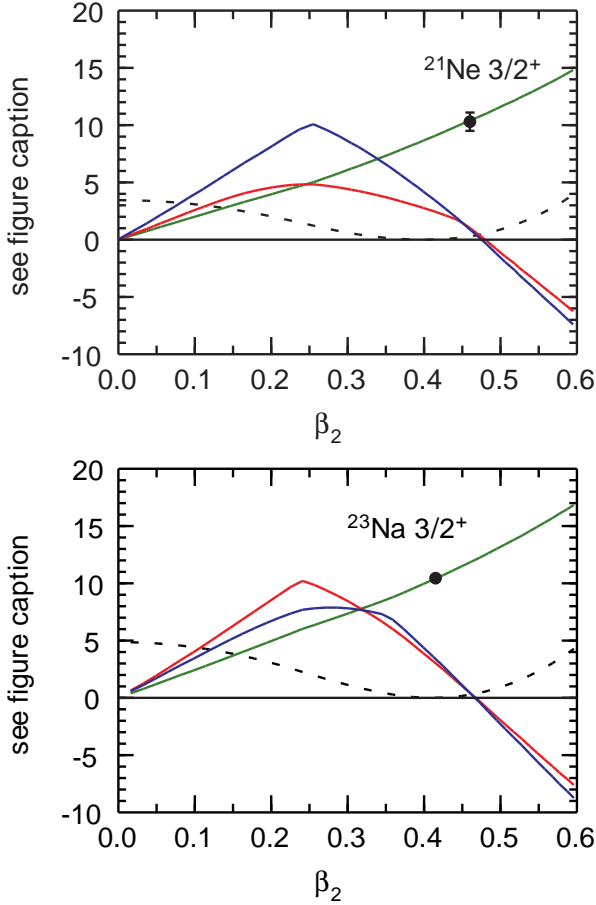}
\caption{Results of the SCMF calculations for $^{21}$Ne and $^{23}$Na
The labels and units are the same as Fig. 1.
}
\end{figure}

The SCMF model has proved to be quite reliable for calculating
matrix elements of one-body operators such as the $  Q  $  in deformed nuclei
\cite{Delaroche}.
Here we use the Hartree-Fock-Bogoliubov method \cite{RS.84}, \cite{Rob.12},
with the Gogny D1S
interaction \cite{D1S}.
The odd-particle orbital is blocked and
time-odd fields are taken into
account in the self-consistent process. Axial symmetry is preserved
so that the different mean field configurations can be labeled with
the $  K  $ quantum number of deformed nuclei. Reflection symmetry is
allowed to be broken, but in the isotopes treated here parity
remains a good quantum number. SCMF results for
$^{21}$Ne, $^{133}$Cs,
$^{173}$Yb and $^{201}$Hg are shown in Figs. 1 and 2 as a function
of the deformation parameter $\beta_{2}$.
For $^{133}$Cs and $^{201}$Hg the energy has a broad minimum around
$\beta_{2}$=0, whereas $^{21}$Ne and $^{173}$Yb have a large prolate deformation.

The experimental $  Q  $ values as shown in these
figures lie near the SCMF energy minimum. The $  M  $
cross zero near where the SCMF energy has a minimum.
These results imply $  \mid M_{p}\mid <10  $ and $  \mid M_{n}\mid <10  $ (in units of $  m  $ MeV).
A more precise limit or a non-zero value might be obtainable
if the calculations were extended beyond mean field to include
fluctuations around the energy minima, for example by
generator coordinate
method (GCM). For $^{133}$Cs, $^{173}$Yb and $^{201}$Hg, these limits are
a factor of five smaller than the values obtained Flambaum et al.
\cite{flam}, \cite{lackenby}, because they do not take into account
the total energy minimization.

The reason for this result
can be seen easily
with a very simple density-functional model which generalizes the
harmonic oscillator model of Bohr and Mottelson \cite{bm}.  We take the
energy functional as
$$
E = \langle \Psi \mid {p^{2}\over 2m}\mid  \Psi  \rangle + \int d^{3} r
{\cal V} [\rho (r)],       \eqno({5})
$$
where $  \cal V  $ is an interaction-energy functional depending only
on the local density
$  \rho(r) = \langle \Psi \mid  a^{\dagger}_{r} a_{r} \mid  \Psi\rangle  $.
Consider the change in energy when the wave function
is changed by the scaling transformation for the $  i  $ nucleons
$  \Psi'(r_{i}) = \Psi(r'_{i})  $ where
$  r' = (x',y',z') = (x\,e^{-\varepsilon/2},y\,e^{-\varepsilon/2},z\,e^{\varepsilon})  $.
The interaction energy remains the same with the new wave function
because the Jacobian for the transformation of variables is unity,
ie. $   d^{3} r = d^{3}r'  $.  The
kinetic term does change, depending on $  \varepsilon  $ as
$$
{ 1\over 2 m}\langle p^{2} \rangle^{\varepsilon} = {1\over 2 m}
\left( \langle p_{x}^{2} \rangle e^{\varepsilon} +
\langle p_{y}^{2} \rangle e^{\varepsilon}
+\langle p_{z}^{2} \rangle e^{-2\varepsilon}  \right).       \eqno({6})
$$
The energy is minimum in the ground state implying
$   d \langle T \rangle^{\varepsilon}/ d \epsilon =0  $.  Carrying out the
algebra, one finds that the derivative vanishes only
if $  M = 2\langle p_{z}^{2} \rangle-\langle p_{x}^{2} \rangle-\langle p_{y}^{2} \rangle=0  $.
As discussed in \cite{bm}, the equilibrium condition is the isotropy of
the velocity distribution; it is also related to the Bohr-van Leeuwen theorem
of absence of magnetization in the equilibrated classical gas of charged
particles.
This result applies to the isoscalar combination
of the $  M  $ values, $  M_{p}+M_{n}  $. It is possible that there is still some
non-zero isovector component to $  M  $ (proportional to $  M_{p}-M_{n}  $). Also the
momentum-dependent part of the interaction could give some non-zero
isoscalar part. There is an implicit momentum
dependence associated with the exchange operators
built into the D1S.  Also, the spin-orbit term
involves the momentum explicitly. But the D1S interaction is
dominated by the central terms.

In \cite{Smiciklas} the matrix element for $^{21}$Ne was obtained from the
simple assumption that it is described by a neutron in the
$  0d_{3/2}  $ orbital outside of a $^{20}$Ne core. The result from this
model is given
in Table I. This simple model
does not reproduce the experimental value for $  Q_{p}  $.
$^{21}$Ne is better described in the
full $  0d_{5/2},1s_{1/2},0d_{3/2}  $ ($  sd  $)
shell-model basis with USDB Hamiltonian that has been globally validated on properties of
nuclei in that mass region \cite{usd}.
For $^{21}$Ne the spin matrix elements assuming the simple $  0d_{3/2}  $
model are $  <S_{zp}>\, = 0  $ and $  <S_{zn}>\, = -0.3  $ and the magnetic
moment is $\mu$=1.148. The full
$  sd  $ CI results are $  <S_{zp}>\, = 0.022  $ and $  <S_{zn}>\, = 0.292  $
and $\mu$ = -0.720. The latter is in reasonable agreement with the
experimental value of $\mu_{exp}$ = -0.662.

The CI results for the quadrupole matrix elements
are given in Table I. The
calculated $  Q_{p}  $ is about a factor
of two smaller than experiment.
It is well known that the
quadrupole observables require an effective charge \cite{usdq}.
This comes from  core polarization that is related to the
admixture of the giant quadrupole resonance at an oscillator
energy of 2$\hbar\omega$.
Thus the quadrupole
moments are calculated as
$$
Q_{p}=Q_{p}^{sd}(1+\delta _{pp})+Q_{n}^{sd}\delta _{np},
$$
and
$$
Q_{n}=Q_{n}^{sd}(1+\delta _{nn})+Q_{p}^{sd}\delta _{pn}.       \eqno({7})
$$
where $  \delta _{vc}  $ are the corrections due to the polarization of the core
nucleons ($  c  $) by the valence nucleons ($  v  $). For $  N\sim Z  $ one can use
$  \delta _{pp}=\delta _{nn}=\delta _{p}  $ and
$  \delta _{pn}=\delta _{np}=\delta_{n}  $. Values of
$  \delta_{p}=0.36  $ and $  \delta_{n}=0.45  $
are the effective
charge parameters appropriate for $  sd  $-shell E2 observables \cite{usdq}.
The results
with these effective charges are in Table I
labeled CI $+$ CP. With effective charges the $  Q_{p}  $ is enhanced and agrees with experiment.
We can also note that the consideration of small deformations in \cite{lackenby} is
not reliable as quantum fluctuations become very large.

The same polarization physics applies for the momentum anisotropy operator,
but with the opposite sign of the effective charge.
The expression is the same as Eqs. (7) but with $  Q  $ replaced by $  M  $
and a change of sign for all of the $\delta$.
The sign may be
seen from the perturbative formula for the polarization contribution
to the moment of an operator $  \cal O  $,
$$
\delta {\cal O} =  \sum_{{}p,h} \langle p\mid  V\mid  h \rangle
{ 1 \over E_{p} - E_{h}}
\langle p\mid  {\cal O}\mid  h \rangle,       \eqno({8})
$$
where $  p,h  $ are particle and hole orbitals.
For harmonic oscillator orbitals $  p  $ and $  h  $ are two major
shells apart ($  \Delta N = 2  $ where $  N=2n+\ell   $), and
the matrix elements of the operators $  \hat Q  $ and $  \hat M  $ are related by
$$
\langle p\mid  \hat Q  \mid  h\rangle =
- {1\over m^{2} \omega_{0}^{2} } \langle p\mid  \hat M \mid  h \rangle.       \eqno({9})
$$
where $  \omega_{0}  $ is the oscillator frequency.
Applying the above effective charges with the opposite sign, we
obtain the $  M  $  given in Table I. The $  M  $ are strongly reduced by core-polarization.
The SCMF results at the energy minimum of Fig. 2 are given in the
last line of Table I. The CI and SCMF results are fairly consistent.
Given this consistency, we suggest that the $  M  $ matrix elements
for $^{21}$Ne are small but not zero; $  M_{p}=3(1)  $ and $  M_{n}=5(2)  $ $  m  $ MeV.

\begin{table}
\caption{Quadrupole matrix elements for $^{21}$Ne, I$^{ \pi }$=3/2$^{ + }$.
CP is the core-polarization correction.
 The experimental
value is from \cite{nnds}.}
\begin{center}
\begin{tabular}{|c||c|c||c|c|}
\hline
  &  $  Q_{p}  $  & $  Q _{n}  $ & $  M_{p}   $ &
$  M_{n}  $  \\
  & fm$  ^{2}  $ & fm$  ^{2}  $ & $  m   $ MeV & $  m   $ MeV
\\
\hline
 experiment  & 10.3(8) &  &  &   \\
   $  \nu  0d_{3/2}  $ & 0 & -4.5 & 0 & -18.2  \\
  CI &  5.4 & 6.4 & 21.9 & 25.9  \\
  CI $+$ CP &  10.2 & 11.0 & 2.7 & 7.0 \\
   SCMF &  8.6 & 9.7 & 2.8 & 4.2 \\
 \hline
\end{tabular}
\end{center}
\end{table}

\begin{table}
\caption{Quadrupole matrix elements for $^{23}$Na, I$^{ \pi }$=3/2$^{ + }$.
CP is the core-polarization correction. The experimental
value is from \cite{nnds}. }
\begin{center}
\begin{tabular}{|c||c|c||c|c|}
\hline
  &  $  Q_{p}  $  & $  Q _{n}  $ & $  M_{p}   $ &
$  M_{n}  $  \\
  & fm$  ^{2}  $ & fm$  ^{2}  $ & $  m   $ MeV & $  m   $ MeV
\\
\hline
 experiment  & 10.45(10)  &  &  &   \\
  CI &  5.8 & 6.3 & 23.7 & 25.2  \\
  CI $+$ CP &  10.7 & 11.2 & 3.6 & 5.9 \\
   SCMF &  10.3 & 11.3 & 3.2 & 3.6 \\
 \hline
\end{tabular}
\end{center}
\end{table}

In summary, we have presented new
calculations for the momentum matrix elements relevant for low-energy tests of
local Lorentz invariance violation involving polarized nuclear spins.
With our self-consistent mean-field (SCMF) calculations
we showed that momentum quadrupoles are small, and explain
this using a variational principle for the energy with
momentum-independent interactions.
Previous calculations by Flambaum \cite{flam}, \cite{lackenby}
make a connection between the experimental spatial quadrupole
moment and the momentum quadrupole moments. In contrast, we
find that these two kinds of moments are not connected, the
spatial matrix element is strongly enhanced in deformed nucleus, but
the momentum matrix element is small and close to zero in
the nuclei studied here.

For the heavy nuclei the best we can do with the SCMF model is to place
an upper limit on the quadrupole momentum matrix element $  M  $
of about 10 $  m  $ MeV.  But the $  M  $ values are not zero,
and until better calculations
can be done, we would suggest a nominal value of one $  m  $ MeV
be used to interpret LLIV experiments for heavy nuclei. Even though this
is much smaller than Flambaum's estimates \cite{flam}, \cite{lackenby},
it still provides useful constraints
on the non-standard model parameters from  LLIV experiments.

The generator coordinate
method often used in nuclear physics
to deal with quantum correlations beyond the mean field require
overlaps of operators between Hartree-Fock-Bogoliubov wave functions.
The explicit form of those overlaps do depend on the quantum numbers of the system,
and time-odd effects have to be considered in odd-A nuclei. As a consequence,
just a very small fraction of the GCM calculations so far have addressed
odd systems (see Ref. \cite{gcm1} for a recent example, that includes symmetry
restorations). Recent advances in the techniques required \cite{gcm2} suggest
that GCM computer codes for odd-A systems will become available
soon and will be as popular as the
ones for even-even systems. As the computational cost of the GCM scales
moderately with mass number A, the new developments will allow calculations
in both light and heavy nuclear systems.

For $^{21}$Ne and $^{23}$Na we use both the
self-consistent mean-field model and configuration interaction models.
The consistency of these models provides some confidence in
a non-zero $  M  $ value that involves both protons and neutrons.
Previously,
a simple model based on a valence neutron was used to
put limit on the LLIV non-standard model parameters for
the neutron \cite{Smiciklas}. Our result implies that this limit should be
applied to a combination of proton and neutron that is
approximately the isoscalar combination of the two.
The $  M  $ moments should be calculated in ab-initio approaches
to light nuclei \cite{ab1}, \cite{ab3}, \cite{ab4}, to check the results obtained
in the more phenomenological models used here.

Acknowledgements: We would like to acknowledge discussions with Michael Hohensee and
Victor Flambaum. BAB and VZ were supported in part by NSF grant PHY-1404442
and by the grant from the Binational Science Foundation US-Israel.
GB was supported in part by U.S. Dept. of Energy grant No.~DE-FG02-00ER41132.
LMR was supported in part by the Spanish grants Nos. FPA2015-65929 MINECO and
FIS2015-63770 MINECO,
and by the Consolider Ingenio 2010 program MULTIDARK CSD2009-00064.
MVR was supported in part by NSF grant PLR-1142032

\end{document}